# Epidemiological study of novel coronavirus (COVID-19)

**Jagadish Kumar**[1]*, **K. P. S. S. Hembram**[2*]

[1]Department of Physics, Utkal University, Bhubaneswar, 751004, India

[2]Independent Researcher

*Corresponding author: jagadish.physics@utkaluniversity.ac.in,

hembramhembram@gmail.com

**ABSTRACT:** We report a statistical analysis of some highly infected countries by the novel coronavirus (COVID-19). The cumulative infected data were fitted with various growth models (*e.g.* Logistic equation, Weibull equation and Hill equation) and obtained the power index of top ten highly infected countries. The newly infected data were fitted with Gaussian distribution with the peak at ~40 days for the countries whose infection curves are seem to be saturated. The similarity in growth kinetics of infected people of different countries provides first-hand guidelines to take proper precautions to minimize human damage.

**1. Introduction**

The outbreak of coronavirus (COVID-19) from Wuhan, China, that causes severe respiratory tract infections in humans has become a global health concern [1,2]. Although the major mode of transmission via respiratory droplets, contaminated hands, surfaces have been found out [3-8], the other modes can't be neglected [9]. The factors like low air temperature and low humidity highly influence the transmission of COVID-19 [10,11]. The population density, qualitative medical care, etc. also affect the control of COVID-19 [12,13]. The outcome of all the factors and modes of transmission lead to the infection in human with suffering in pain and to the death in one hand. On the other hand, although there is no direct vaccine available to cure it, alternative medicines are used to treat the patients to recover [1,2].

Till now (end of March, 2020) the epidemic is under mostly control only in China and Republic of Korea (see Fig. 3(a,b)), whereas it is still not controlled in most countries, despite their attempts in various modes. There are attempts to understand the behaviour of the epidemic in different mathematical formalism, they lack many details in some factors or are restricted to a particular locality [14-24].

Based on the WHO data (till date), we investigate the behaviour of COVID-19 spread for various countries to take precautions. This can provide first-hand guidelines for strategically controlling the epidemic.

**2. Methods**

*2.1 Data*

We have taken the data from World Health Organization website (WHO) for all the COVID-19 infected countries [1], although individual affected countries have their individual database. However, we prefer the data from "WHO" for convenience.

## 2.2 Statistical analysis

The cumulative infected data were analyzed based on various growth models (*e.g.* Logistic equation, Weibull equation and Hill equation) to obtain the power index. The newly infected data were fitted with Gaussian distribution function.

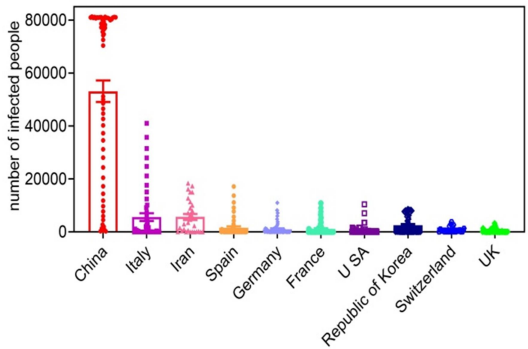

**Fig. 1.** Cumulative infected data from the top ten highly infected countries.

## 3. Results

The COVID-19 is spreading rapidly all over the world and WHO declared it as "*public health emergency of international concern (PHEIC)*". Globally there are 292142 confirmed cases and death is 12784 till the date March 21, 2020 [1]. The cumulative data of top ten highly infected countries data are shown in Fig. 1. The flatness of the top data of any country represents the saturation of new infection (*e.g.* China). The other behavior represents the increase in the number of infected people with time. Initially the infection rate may be low but increases as time progress and reach to a saturation value. For the comparison of infection rate, normalized curves are plotted for these countries (see Fig. 2).

Further, the infection curves are fitted with the Hill equation, *i.e.* $Y = \frac{AT^m}{K+T^m}$, where '$Y$' is the cumulative data of infection, '$T$' is the time, $m$ is the power index, '$A$' and '$K$' are the constants. It is observed that the data is saturated for the two countries *i.e.* China and Republic of Korea (Fig. 3 (a, b)). This may be due to various factors developed in the later stage like good medical facilities with medications and awareness. It is revealed that for China and Korea the saturation is observed with less newly affected persons. But for other countries the infection trend is increasing. Also it is revealed that the nature of increase trend is similar in nature. Although little discrepancies are observed among various countries, it can be attributed to the local natural factors like temperature, humidity and human factors like GDP, population density etc. )

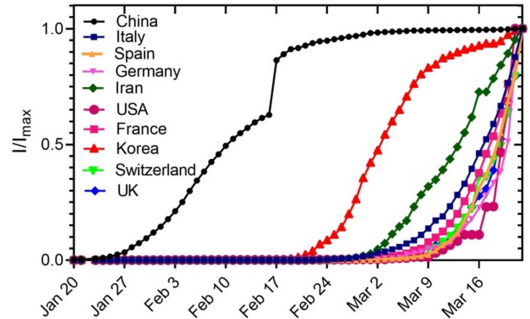

**Fig. 2.** Normalized cumulative data of top ten highly infected countries.

We have also fitted the cumulative infection curve with Logistic and Weibull equations to compare the results with Hill equation and calculated the power index (*m*) which gives the information about the infection rate. We observed that the values are very close to each other for all the models (see Fig. 4). The power index value ranges between 3.75 - 32.23 for the top ten highly infected countries. As the cumulative infection curve of China and Republic of Korea are almost saturated, one can compare the exponent value of these two countries. The exponent is low (3.75) for China whereas for the Republic of Korea it is 12.01. The small exponent indicates the low infection rate *i.e.* the infection rate of China is relatively small as compared to Republic of Korea. For other countries the cumulative data is not reached to the saturation value, so the interpretation of the power index may be unrealistic. For example, in the case of the USA the exponent is quite high because there is a sharp increase of infection at this stage but it may slow down due to prevention and medical facility as time progresses.

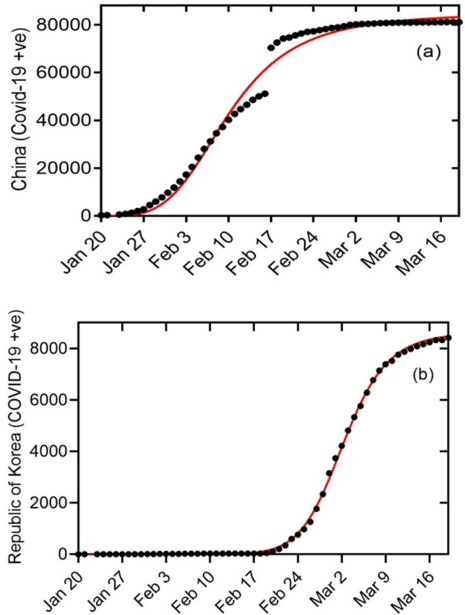

**Fig. 3**. Cumulative infection curve fitted with Hill equation for (a) China and (b) Republic of Korea.

Due to different transmission mode of COVID-19 spreading, many people are getting infected on daily basis although there are different precautions are available. The daily infected people of top ten countries are shown in Fig. 5. The mean of newly infected people is quite high for most of the countries.

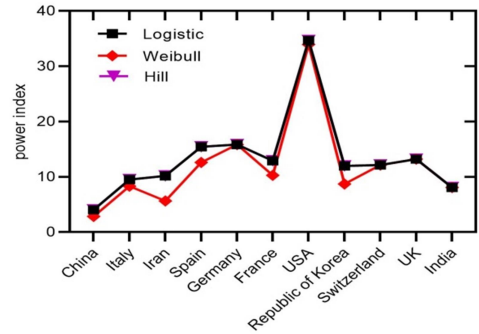

**Fig. 4.** Power index obtained from different model for top ten highly infected countries.

For better understanding, we have also analyzed the daily infection and that is fitted with a Gaussian distribution. While the data from China and Korea are well fit (see Fig. 6 (a,b)), the data from Iran and Switzerland show the change in trend (may be achieving its peak) (Fig. 6 (c,d)). But the data from the countries like Italy and USA shows the increasing in trend (Fig. 6 (e, f)). The average mean are ~35 and ~40 days for China and Republic of Korea respectively. Although many factors (temperature, humidity, GDP, medical facilities, etc.) in other countries are different than that of China and Republic of Korea, they are trying best to control it. Irrespective of that the rate of infection is increasing till now. The rest of the countries data are shown in the Supporting Information.

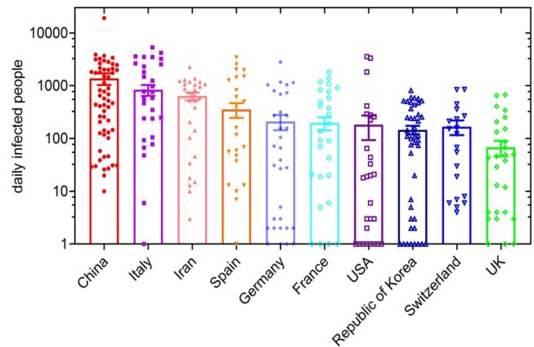

**Fig. 5.** Daily infected data from the top ten highly infected countries.

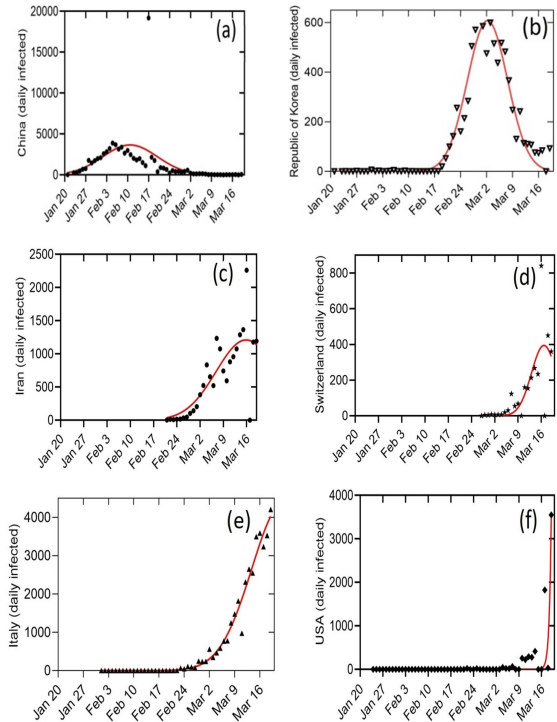

**Fig. 6.** Distribution of daily infected patients fitted with Gaussian function for (a) China, (b) Republic of Korea, (c) Iran, (d) Switzerland, (e) Italy and (f) USA.

The global infected data of COVID-19 shows a power law behavior with exponent 1.7 (see Fig. 7). Although the spreading of coronavirus is different in the respective countries, they show a scale invariance behavior.

**Conclusion**

The present work discussed the statistical analysis of COVID-19 infection data. The cumulative infection of top ten highly infected countries data are fitted with different equations and the power index are estimated. There is a similarity in the growth kinetics of infected people although the rate of infections is different due to various reasons. The infection curve of China and Republic of Korea has almost reached to its saturation value because of various reasons, for example medical facilities, prevention and public awareness etc. Further, the distribution of daily infected people is well fitted with Gaussian function. The cumulative infection data shows the scale invariance property. All the statistical analysis will provides first-hand guidelines to take proper precautions to minimize the infection rate.

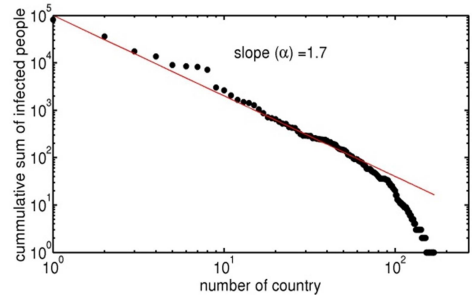

**Fig. 7.** Power law behavior of infection spread.

**Acknowledgements**

We acknowledge World Health Organization (WHO) for the COVID-19 infection data.

**Competing interests**
The authors declare no competing interests.
**Additional information**

*Supporting information* is available at the later part of this manuscript.

**Supporting information for**

# Epidemiological study of novel coronavirus (COVID-19)

**Jagadish Kumar**[1]*, **K. P. S. S. Hembram**[2]*

[1]Department of Physics, Utkal University, Bhubaneswar, 751004, India

[2]Independent Researcher

*Corresponding author: jagadish.physics@utkaluniversity.ac.in,

hembramhembram@gmail.com
The cumulative and daily infected patients from all the infected countries are fitted with Hill equation and Gaussian are shown in the left and right panel respectively.

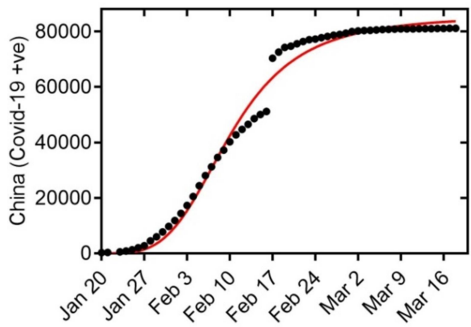 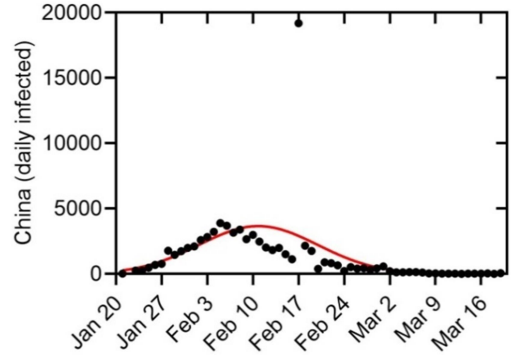

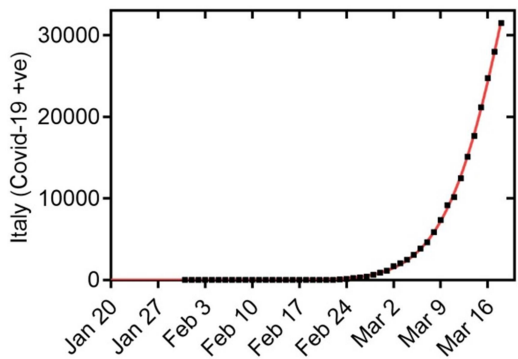 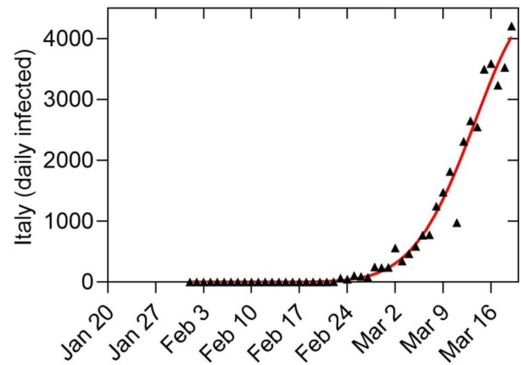

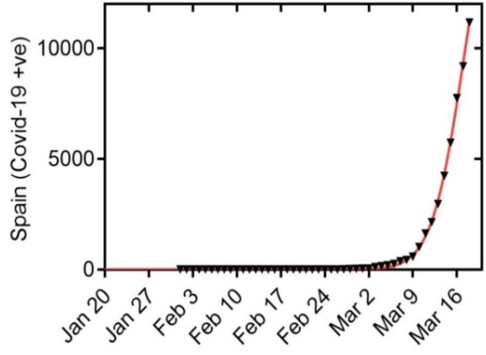
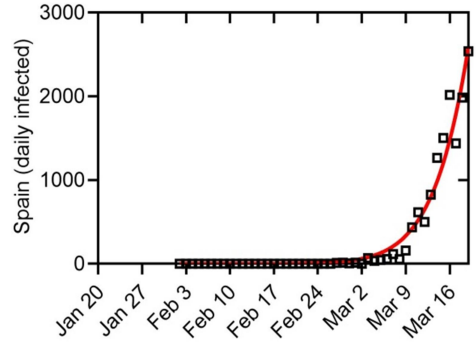
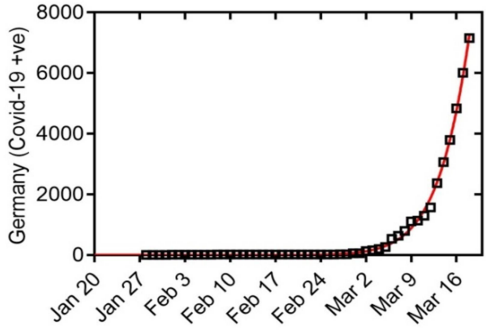
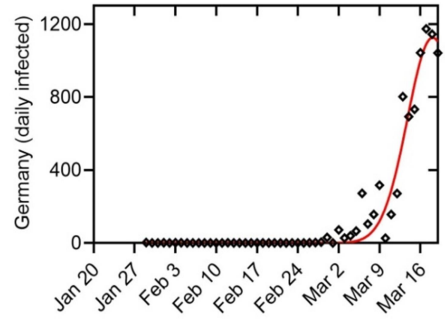
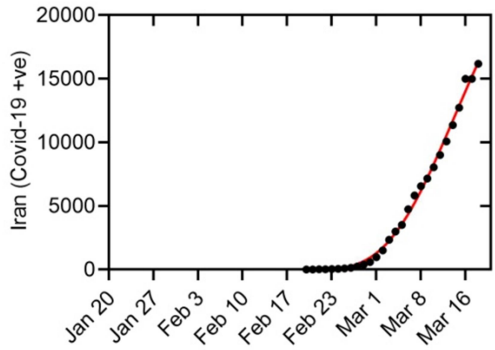
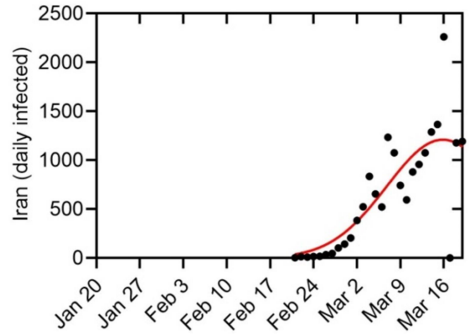
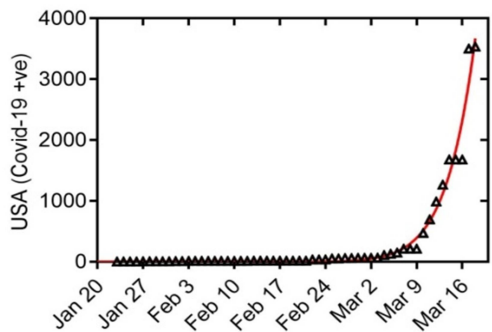
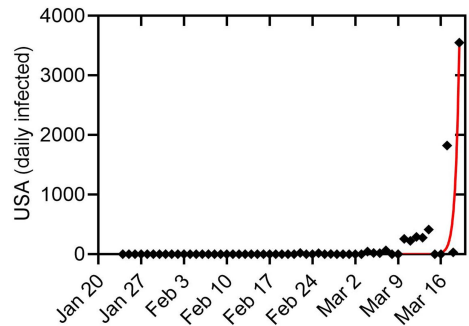

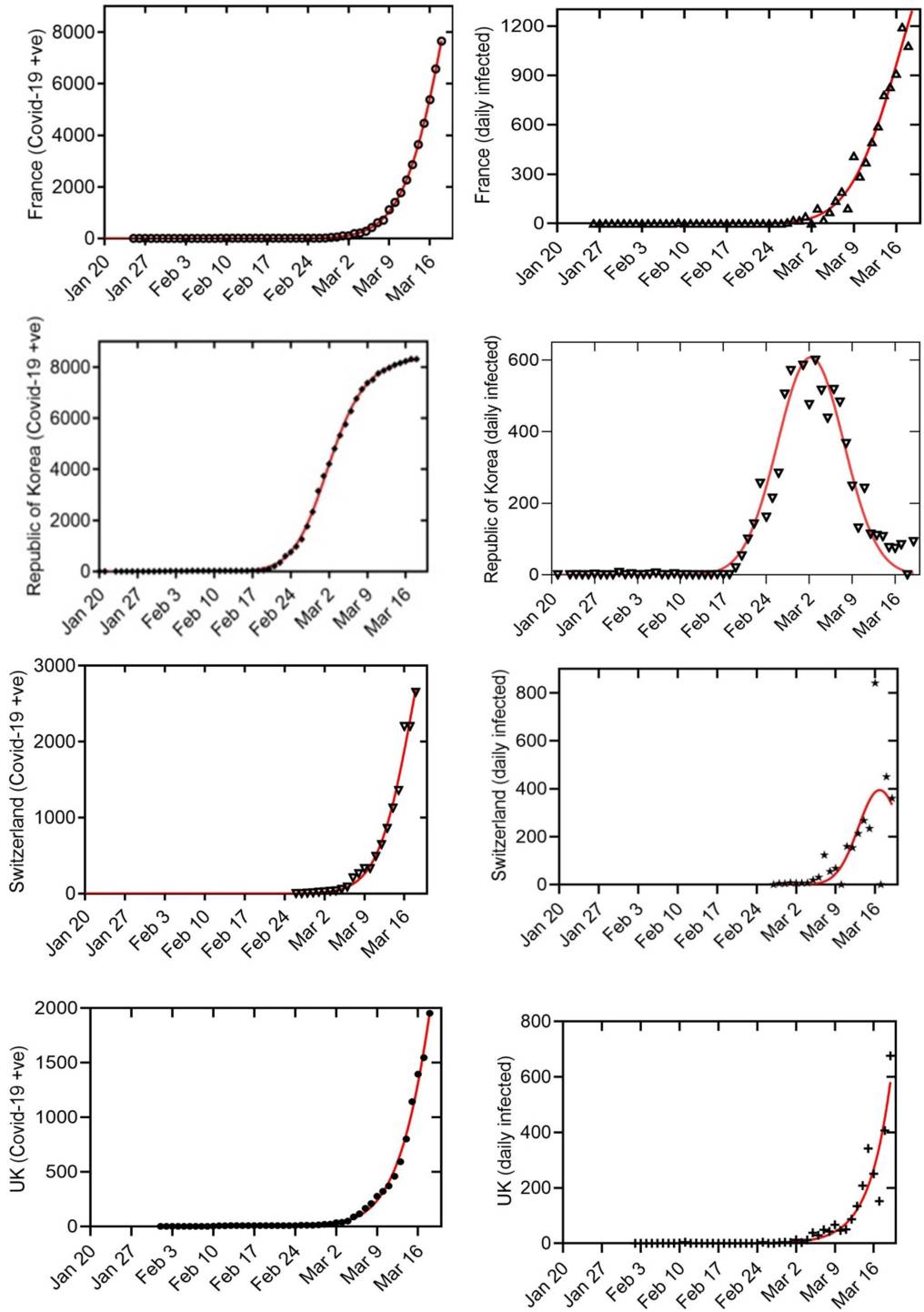

**Fig. SI. 1.** Cumulative (left panel fitted with Hill equation) and daily (right panel fitted with Gaussian distribution) infected patient data from top ten highly infected countries.